\documentclass[traditabstract]{aa}

\usepackage[dvips]{graphicx}
\usepackage{amsmath}

\usepackage{natbib}
\usepackage{txfonts}

\usepackage{amssymb}

\usepackage{subfig}

\newcommand{\comment}[1]{}

\begin{document}

\title{Modeled IBEX/INCA skymaps including the keV-ENA source extinction in the outer heliosphere}

\author{M. Siewert \and H.-J. Fahr}

\institute{Argelander Institut f\"ur Astronomie der Universit\"at Bonn,
Abteilung f. Astrophysik und Extraterrestrische Forschung,
Auf dem Huegel 71, 53121 Bonn (Germany)}

\abstract{Understanding the outer heliospheric interface is a
major challenge, especially in the light of recent observations by
the IBEX and Voyager missions. We present further details on a new
required theoretical correction that has been identified as substantial
in a recent paper, the
so-called source depletion effect. These results complement and expand earlier
calculations of transit-time delays by presenting global skymaps
of Energetic Neutral Atoms (ENAs) calculated with the depletion correction,
comparing them with skymaps calculated without these corrections.
Our results demonstrate that the depletion correction is critical for
interpreting IBEX-High ENA fluxes generated in the inner heliosheath,
and that any attempt to reconstruct the shape of the heliospheric boundary
needs to include the depletion correction, unless arriving at considerably
erroneous results.
}


\maketitle

\section{Introduction}

In the past 10 years, new observations by the two Voyager spacecraft
\citep{fisk05-voy1,jokipii08-voy2,webber-mcdonald-13-voyager-heliocliff}
and the IBEX mission
\citep{mccomas-ibex-09,mccomas-12-ibex-update,mccomas12-ibex-no-bow-shock}
have generated renewed interest on the size
and structure of the heliospheric boundary regions. Particularly ENA (energetic
neutral atoms) skymaps, as produced by the IBEX mission, provide a line-of-sight
integrated intensity that, on the basis of
present understanding, turned out unexpectedly difficult to interpret
\citep{mccomas-14-ibex-models}; in fact, even the spatial region in the
outer heliosphere where the observed ENA fluxes are mainly generated remains unclear till now.

One additional irritation in interpreting ENA data is introduced by the INCA/CASSINI
instrument, a detector observing a structure in the global ENA fluxes between 5 and 50~keV,
i.e. beyond the upper energy limit accessible to IBEX, that is topologically similar
to the IBEX ribbon \citep{krimigis-cassini-belt-09},
however with two subtle differences. First, the INCA ENA data presents a
closed ring-like region of enhanced ENA fluxes, while IBEX observes a region of enhanced
ENA production that does not close upon iself. The second main difference between the
observations made by both missions is the orientation of the feature; the INCA ring
does not perfectly match the placement of the IBEX ribbon. Because of these differences, there
is no consensus on whether both detectors are observing the same ENA source regions.

Aiming to derive a quantitative criterion that might allow to differentiate between
individual models, we have recently identified a needed new correction to the calculations
\citep{sf13-transit-time} that was seemingly absent from all previous calculations.
In the context of ENA production and observation, there are two nontrivial corrections
related to the removal of particles from an initial distribution function that need to
be understood for modeling and interpretation of ENA flux data. The first contribution
was studied by \cite{heerikhuisen08-kappa}
who calculated removal of ENAs along the line-of-sight between the source region and the
detector. ENAs have a mean free path of o(100~AU) with respect to reionisation
processes in the inner heliosphere, and therefore, a significant fraction of ENAs
generated in the heliospheric boundary regions will not reach the detector.

In this study, we calculate for the first time a quantitative estimate of the
other important removal process involved in the ENA production.
In our model, we consider energetic protons that
have been accelerated as pick-up ions at the solar wind termination shock (TS), while
downstream of the TS, in the inner heliosheath, there are no additional heating processes operating,
i.e. there are no further sources of energetic protons. Therefore, keV-ENA production along the plasma
flow lines in the inner heliosheath will automatically break down when the initial
resources of keV-energetic proton ``fuel'' have been extinguished, which limits relevant
line-of-sight contributions to lengths of about 100~AU. In \cite{sf13-transit-time},
we have calculated source depletion factors along a few selected lines-of-sight;
in this publication, we extend these calculations to full global sky maps and compare
the results to earlier calculations where the depletion correction was missing
\citep{fs11-enas-acr,sf11-enas-pui,sf13-enas-spectra}.

\section{ENA skymaps with and without the depletion correction}

In the following, we will not repeat many details of the calculation of the
line-of-sight integral for the resulting fluxes,
the plasma flowline profile and related problems, which
have been discussed earlier in great detail by e.g. \cite{sf13-enas-spectra}.

\begin{figure}
\includegraphics[width=0.6\columnwidth]{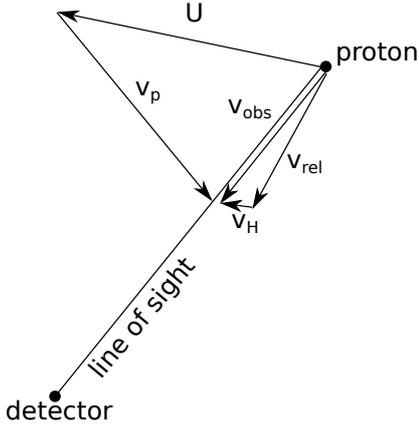}
\caption{Velocity vectors appearing in the ENA charge exchange problem.}
\label{fig-velocities}
\end{figure}

In this study, we calculate ENA skymaps according to the known relation
\citep{lee-et-al-ibex-theory-09}
\begin{equation}
\Phi_{ENA}(v_{obs}, \alpha ,\delta)
 = \int_{r_0}^{r_1} dr \cdot j_{p}(E_{ENA},\vec{r}) \sigma_{ex}(v_{rel}) n_{H},
\label{eq-ena-flux}
\end{equation}
where $\sigma_{ex}$ is the charge exchange crossection, $n_H \simeq 0.1 \, \mathrm{cm}^{-3}$
is the local interstellar hydrogen density, and $j_p(E_{ENA},\vec{r})$ is the source flux
of energetic protons at a point $\vec{r}$. The parameters $v_{ENA}$ and $E_{ENA}$ denote the
energy and velocity of the ions undergoing charge exchange in the rest frame convected
along with the plasma flow, which is related to the ENA velocities
in the observers rest frame ($v_{obs}$) by the relation
\begin{equation}
\vec{v}_{obs} = \vec{v}_{ENA} + \vec{U},
\end{equation}
where $\vec{U}$ is the bulk velocity of the plasma along the flowline and
$\vec{v}_{obs} \parallel \vec{r}_{line-of-sight}$. Obviously, due to the presence
of curved streamlines in the outer heliosheath, this relation introduces a nonlinear
nontrivial correction to the
flow of energetic protons.
Finally, there is the relative velocity $v_{rel}$ between the energetic
plasma proton and the cold hydrogen undergoing charge exchange, for which the most general
definition is given by
\begin{equation}
\vec{v}_{rel} = \vec{v}_{p} + \vec{U} - \vec{v}_H.
\label{eq-def-vrel}
\end{equation}
In the outer heliosphere, and for keV-energetic protons, one finds typical
values of $v_p > U \gg v_H$, i.e. it is possible to adopt the simplification:
\begin{equation}
v_{rel} \simeq v_p + U \simeq v_p.
\label{eq-vrel}
\end{equation}
In this following study, we make use of the first approximation, where $U$ is
the projection of $\vec{U}$ on $\vec{v}_p$.
We also need to mention that we are working in a
pitchangle-isotropic limit, i.e. where the ion distribution function
does not generate any significant anisotropy while moving along the streamlines.
Depending on the energy behaviour of the charge exchange
cross section and the individual velocities involved in the production
process (see Fig. \ref{fig-velocities}), this may not be the case when the
source plasma is propagating along the flowlines; in this study,
we ignore details related to pitchangle-anisotropic terms during this transport.

Finally, the differential
plasma flow $j_p(E_{ena})$ appearing in Eq. \ref{eq-ena-flux}
is defined by \citep[see e.g.][]{lee-et-al-ibex-theory-09}
\begin{equation}
j(E) = \frac{2}{m^2} E f(E) = v f(v) v^2 \frac{dv}{dE} = \frac{1}{m} f(v) v^2.
\label{eq-j-rel}
\end{equation}
Now, we need to introduce a factor for the source depletion described in the introduction.
Therefore, we introduce an additional factor describing the modification of the source
proton distribution function,
\begin{equation}
f(v) \rightarrow f_0(v) \cdot \epsilon(\phi, \theta, r, v),
\end{equation}
where $0 < \epsilon \leq 1$, and $f_0$ is the distribution function
on the immediate downstream side of the termination shock, i.e. before energetic protons
are removed by charge exchange. To calculate this parameter, one
has to integrate the following differential equation along a streamline
from the TS ($\epsilon = 1$, $s_0 = 0$) to the point of interest $s = s(\phi, \theta, r)$
along a line of sight:
\begin{equation}
U \frac{d\epsilon(v,s)}{ds} = -n_H v_{rel} \sigma_{ex}(v_{rel}) \epsilon(s),
\label{eq-depsilon}
\end{equation}
where $ds$ is the line element on the curved streamline.
In principle, a more detailed and complicated transport equation is required,
including terms for proton sources and energy diffusion; however, in the energy
region between 1 and 10~keV, these other terms can be safely ignored.

A formal solution of this differential equation is given by
\begin{equation}
\epsilon(v,s) = n_H \int_{s_0}^s \frac{v_{rel}}{U} \sigma(v_{rel}) ds'.
\end{equation}
Now, it follows from Eq. \ref{eq-vrel} that a full solution of this
equation requires detailed knowledge of the plasma flow velocity field
$\vec{U}(\phi, \theta, r)$ in the outer heliosphere,
which must be extracted from analytical or numerical model calculations.
Unfortunately, most numerical models do not provide this information in an
easily accessible way, so we apply instead the analytical model by
\citep{fahr-fichtner-91-helio},
\begin{equation}
\vec{U} = -\vec{\nabla}_{\vec{r}} \Phi(\vec{r}),
\end{equation}
where the velocity field is obtained from the spatial gradient of a
corresponding velocity potential $\Phi$.
Even though it is possible to calculate the velocity field analytically,
the resulting expressions are very lengthy and unintuitive, and therefore,
we prefer to evaluate this gradient numerically.

\begin{figure*}[tbp]
\begin{center}
\subfloat[$\varepsilon = 0.25$, no depletion correction] {
 \includegraphics[width=0.45\textwidth]{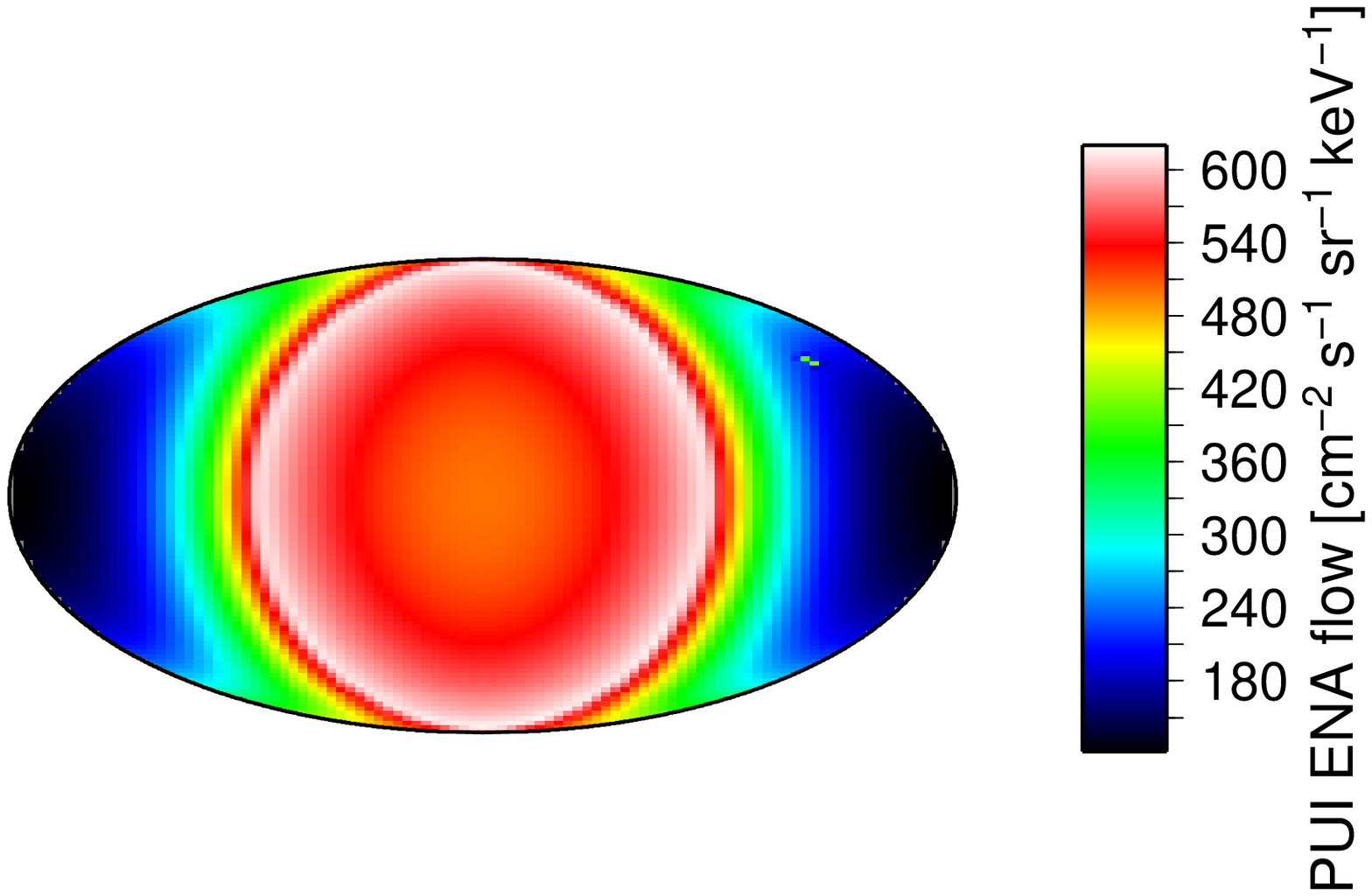}
} 
\hspace{0.05\textwidth}
\subfloat[$\varepsilon = 0.25$] {
 \includegraphics[width=0.45\textwidth]{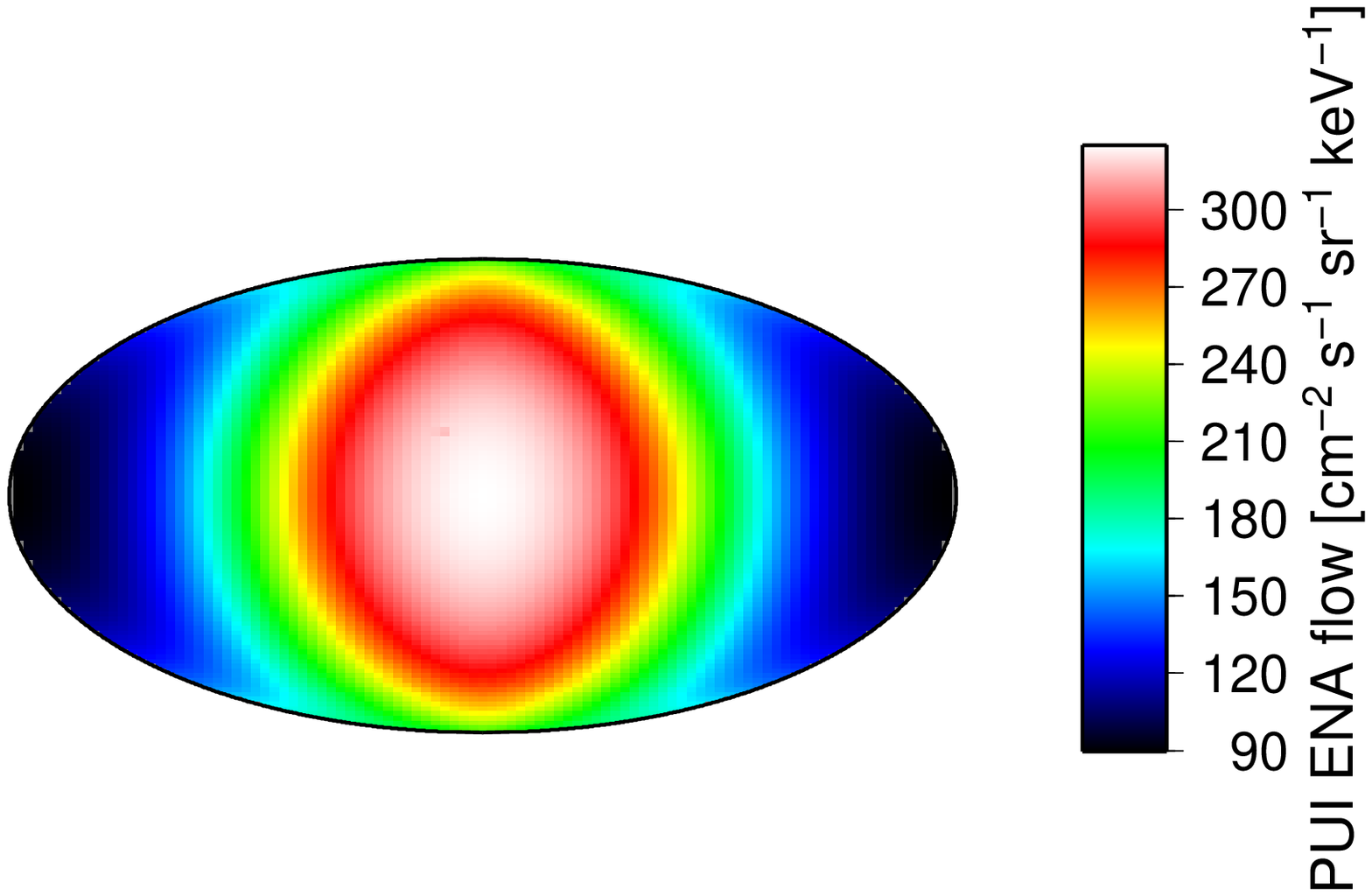}
}
\\
\vspace{-4px}
\subfloat[$\varepsilon = 0.15$, no depletion correction] {
 \includegraphics[width=0.45\textwidth]{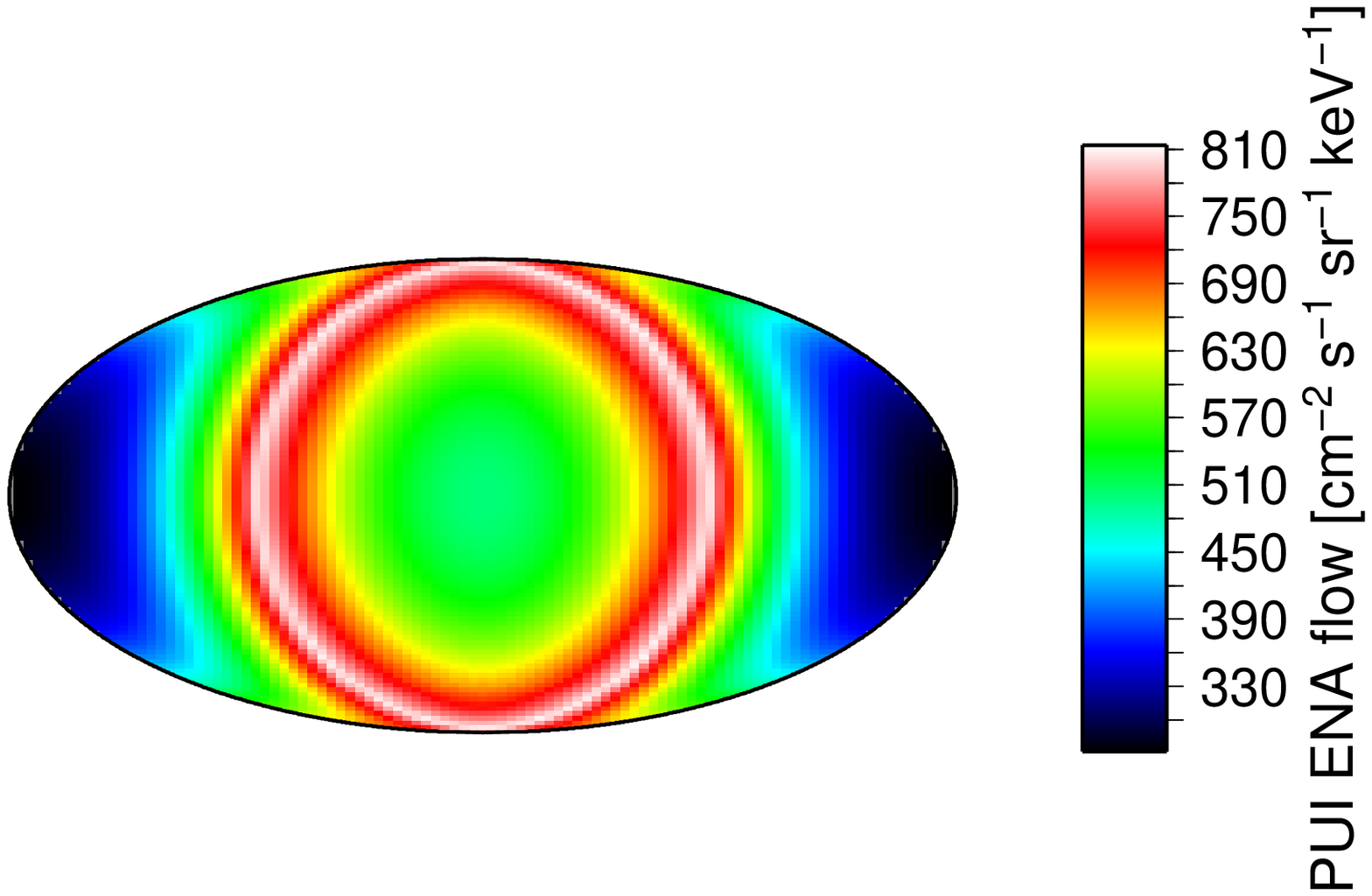}
} 
\hspace{0.05\textwidth}
\subfloat[$\varepsilon = 0.15$] {
 \includegraphics[width=0.45\textwidth]{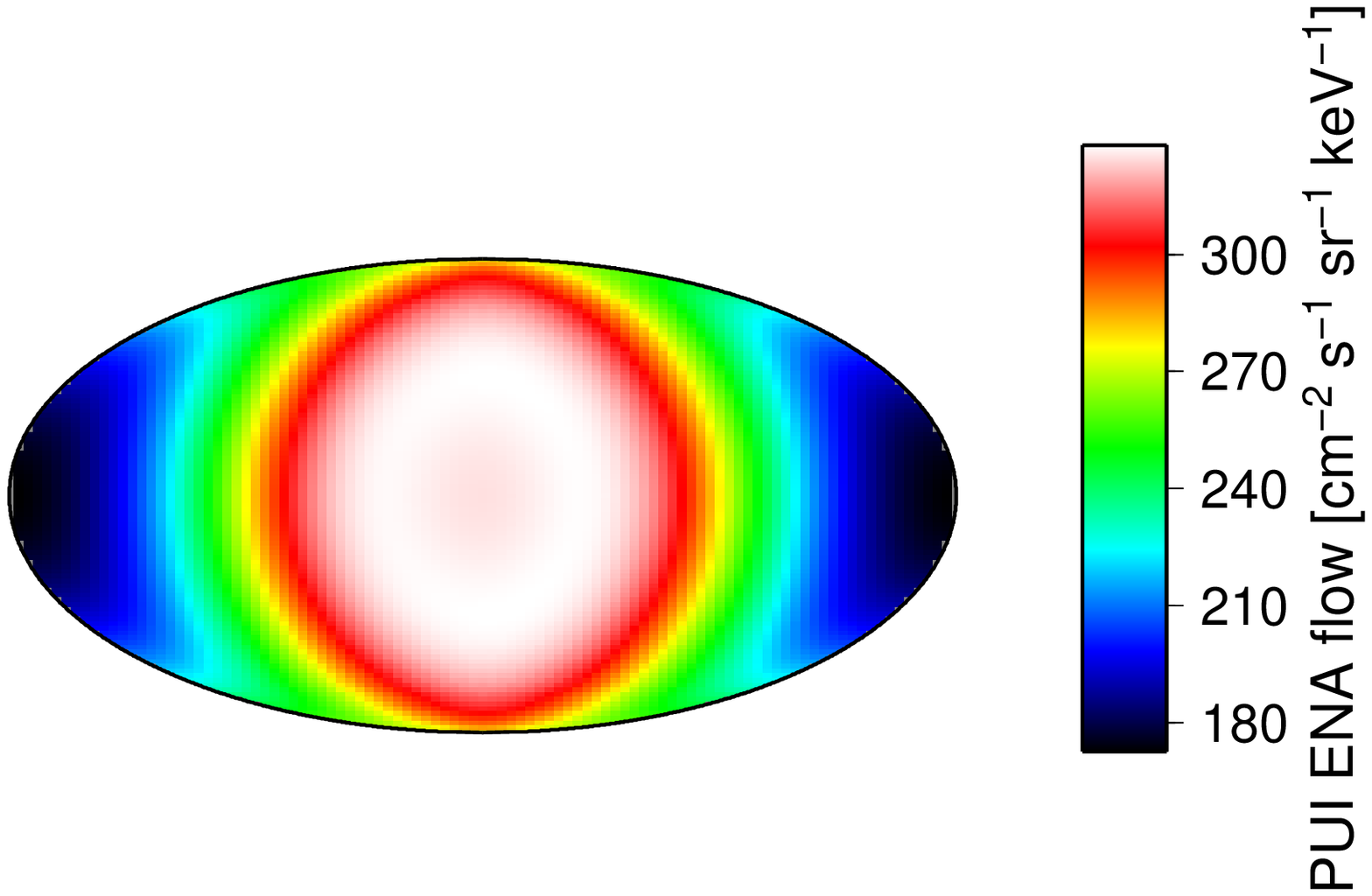}
}
\\
\vspace{-4px}
\subfloat[$\varepsilon = 0.05$, no depletion correction] {
 \includegraphics[width=0.45\textwidth]{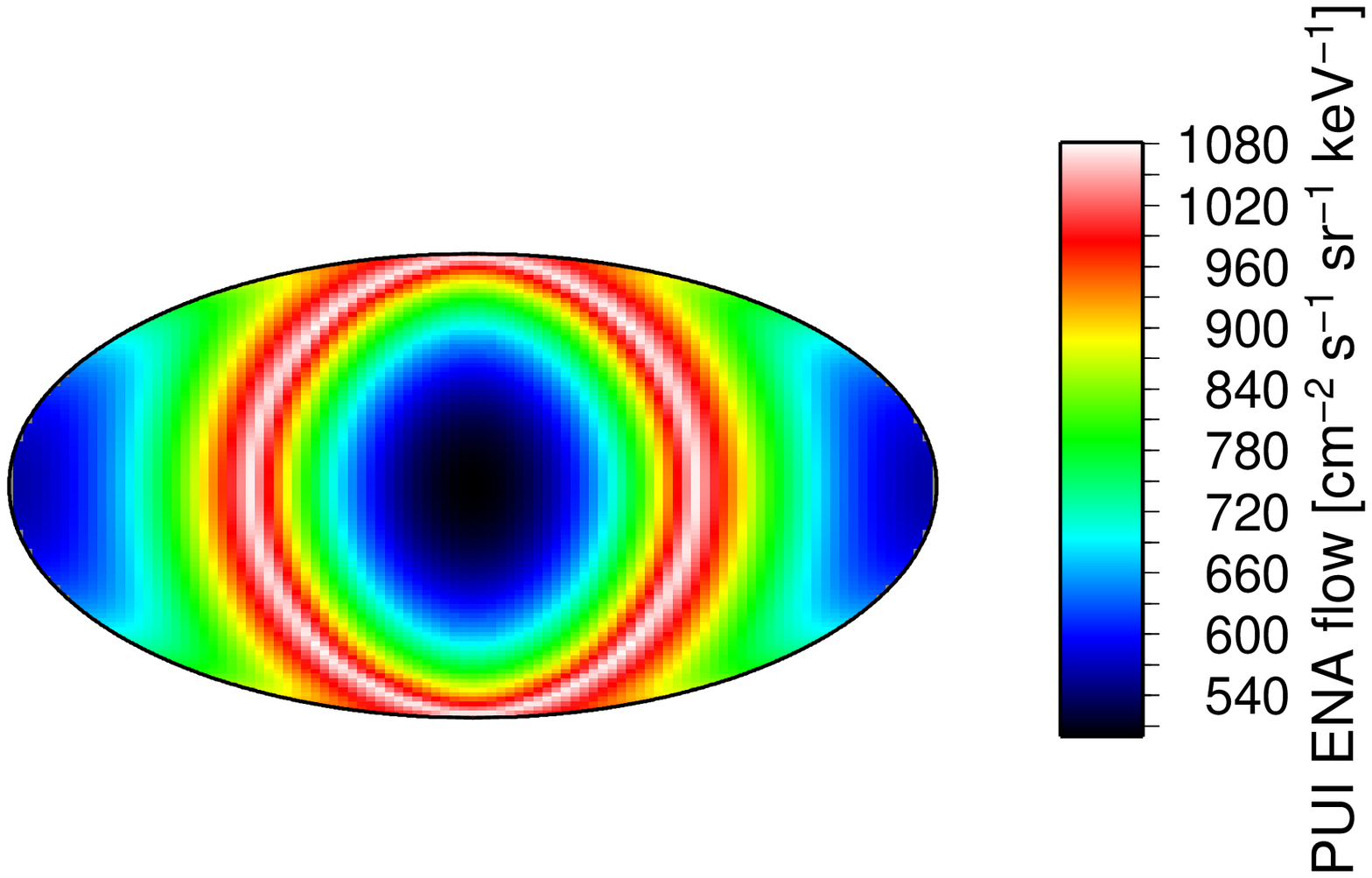}
} 
\hspace{0.05\textwidth}
\subfloat[$\varepsilon = 0.05$] {
 \includegraphics[width=0.45\textwidth]{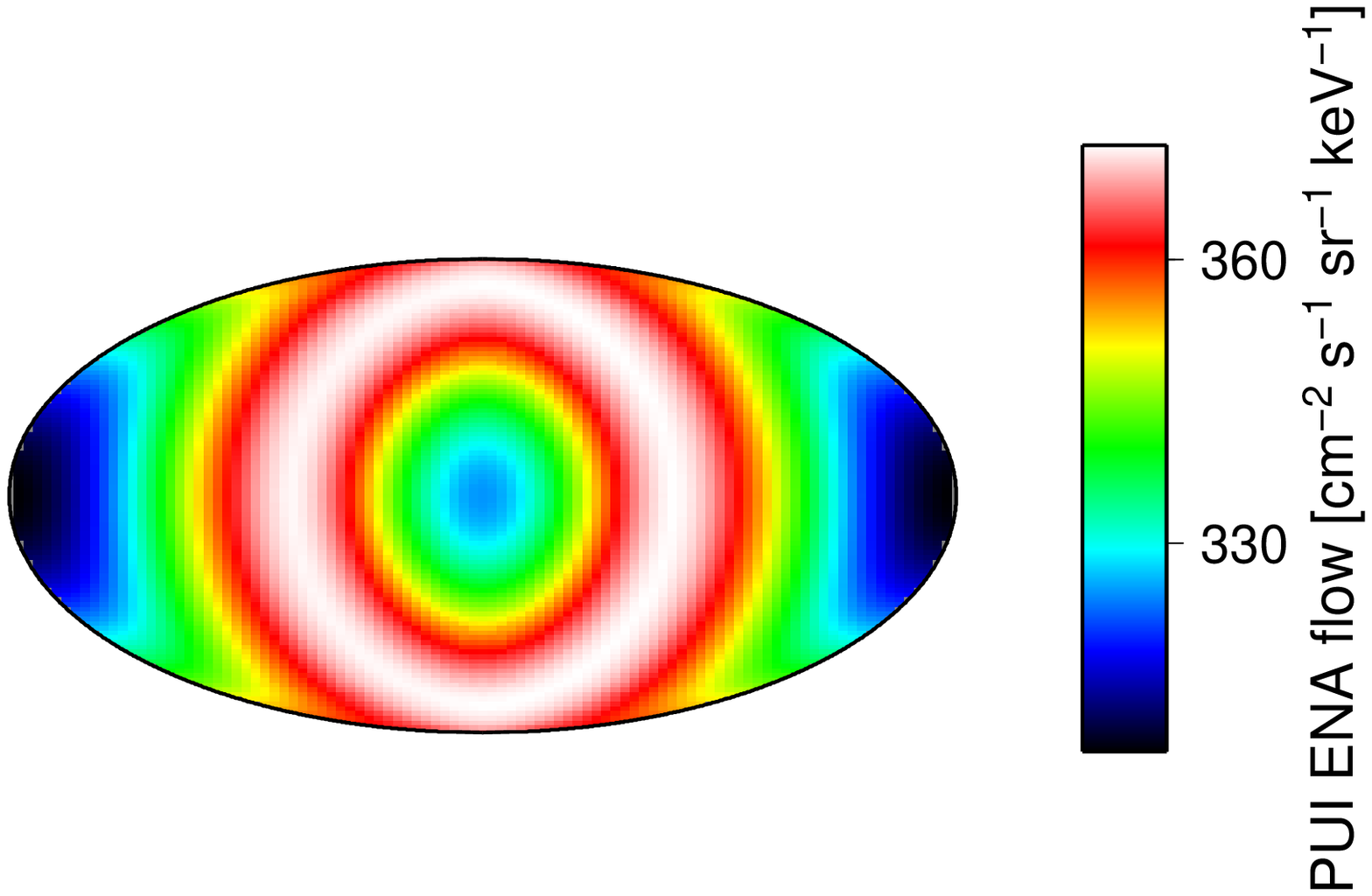}
}
\\
\vspace{-4px}
\subfloat[$\varepsilon = 0.01$, no depletion correction] {
 \includegraphics[width=0.45\textwidth]{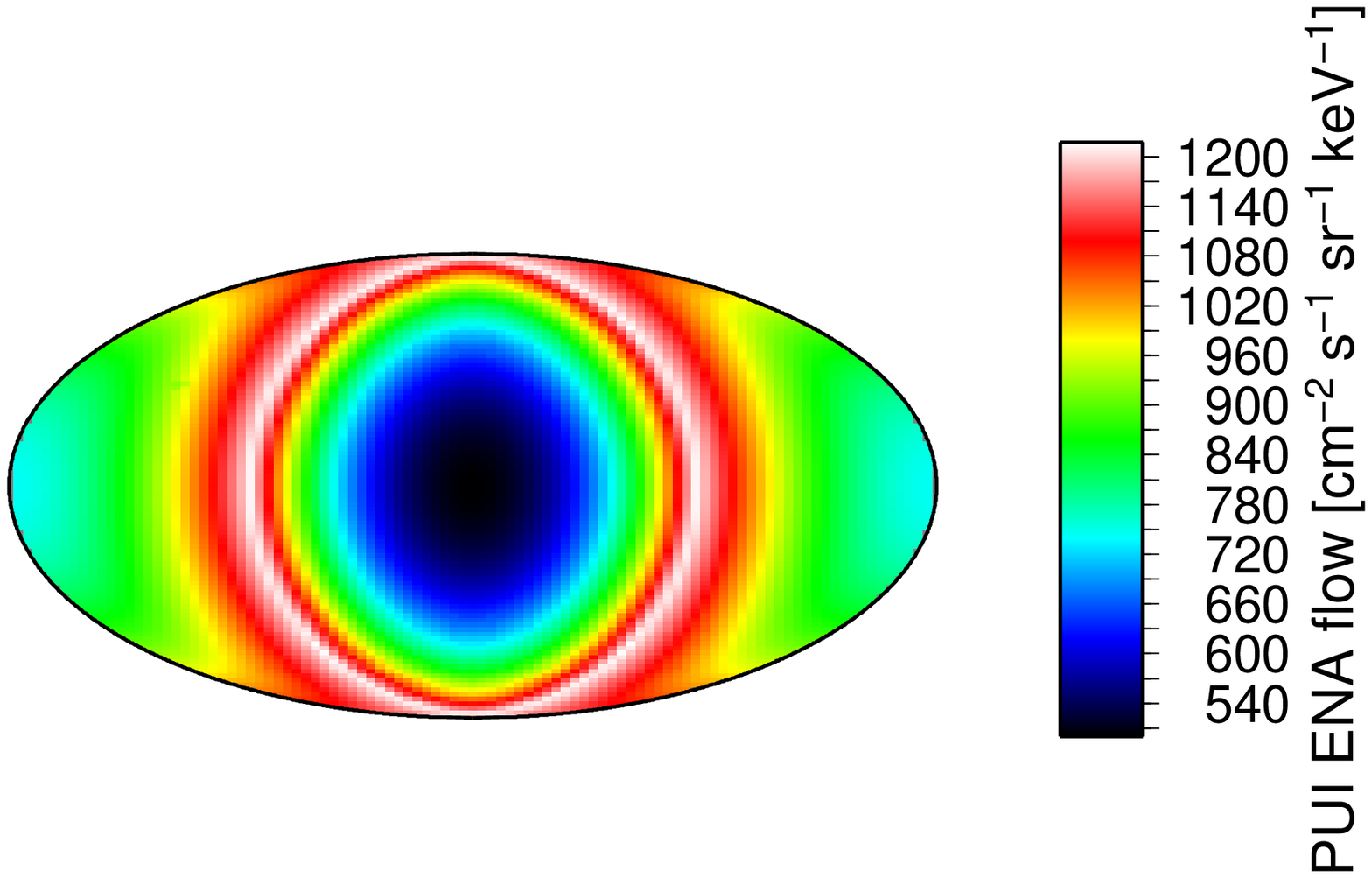}
} 
\hspace{0.05\textwidth}
\subfloat[$\varepsilon = 0.01$] {
 \includegraphics[width=0.45\textwidth]{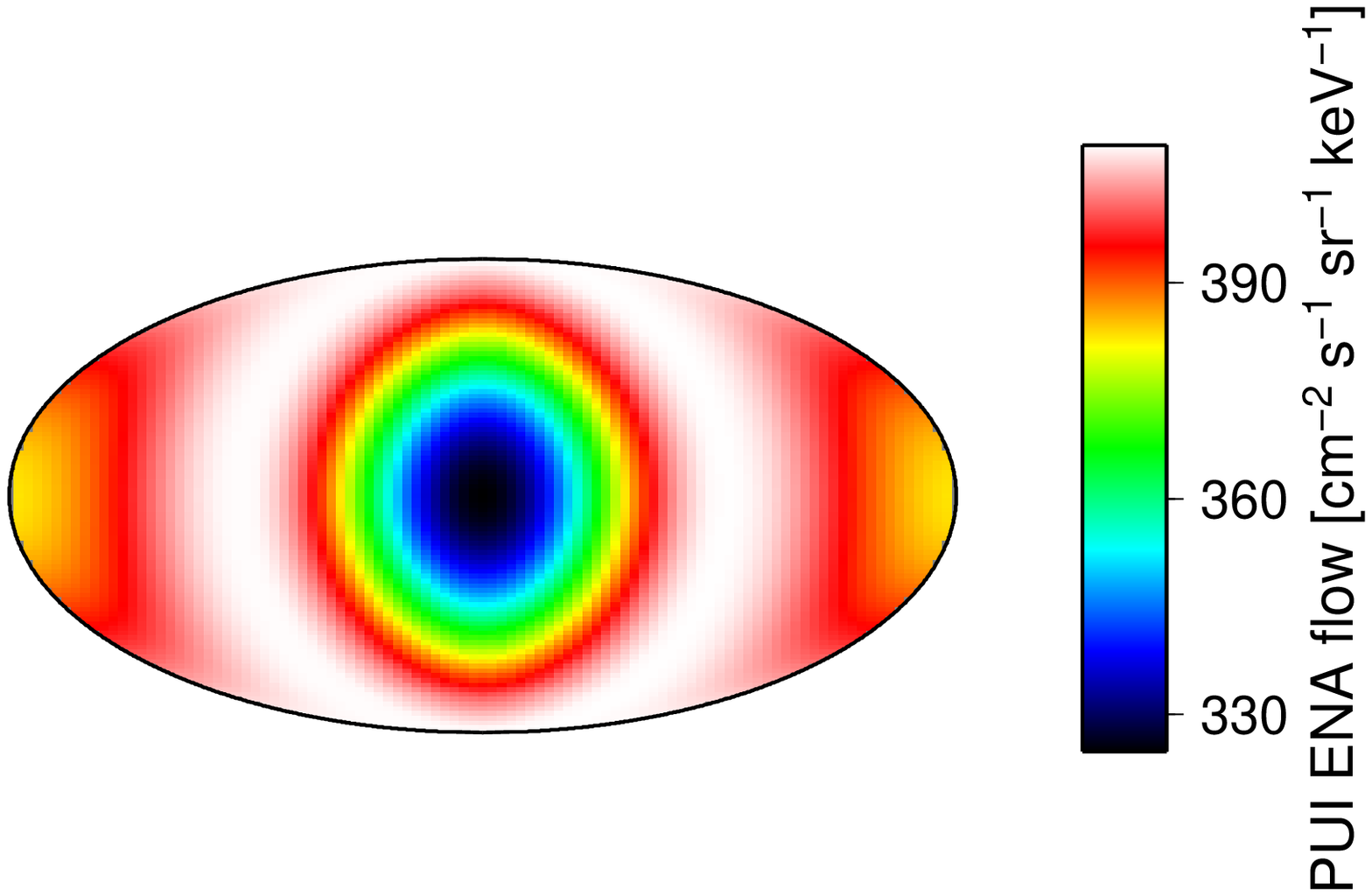}
}
\end{center}
\vspace{-10px}
\caption{ENA flow skymaps at $E_{obs}$ = 1~keV for different TS geometries and the presence/absence of the depletion correction.}
\label{fig-ena-skymap}
\end{figure*}

\begin{figure*}[tbp]
\begin{center}
\subfloat[$\varepsilon = 0.05$] {
 \includegraphics[width=0.45\textwidth]{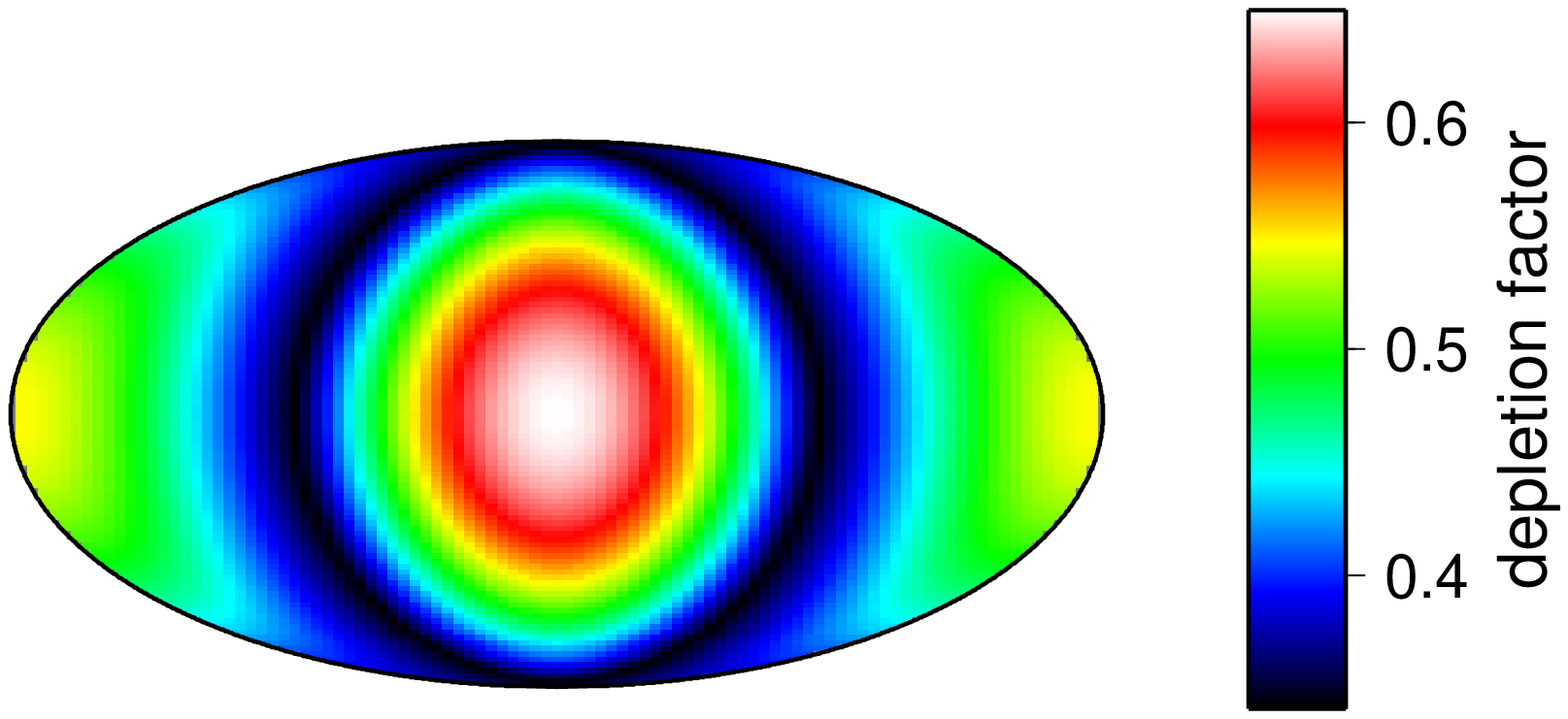}
} 
\hspace{0.05\textwidth}
\subfloat[$\varepsilon = 0.01$] {
 \includegraphics[width=0.45\textwidth]{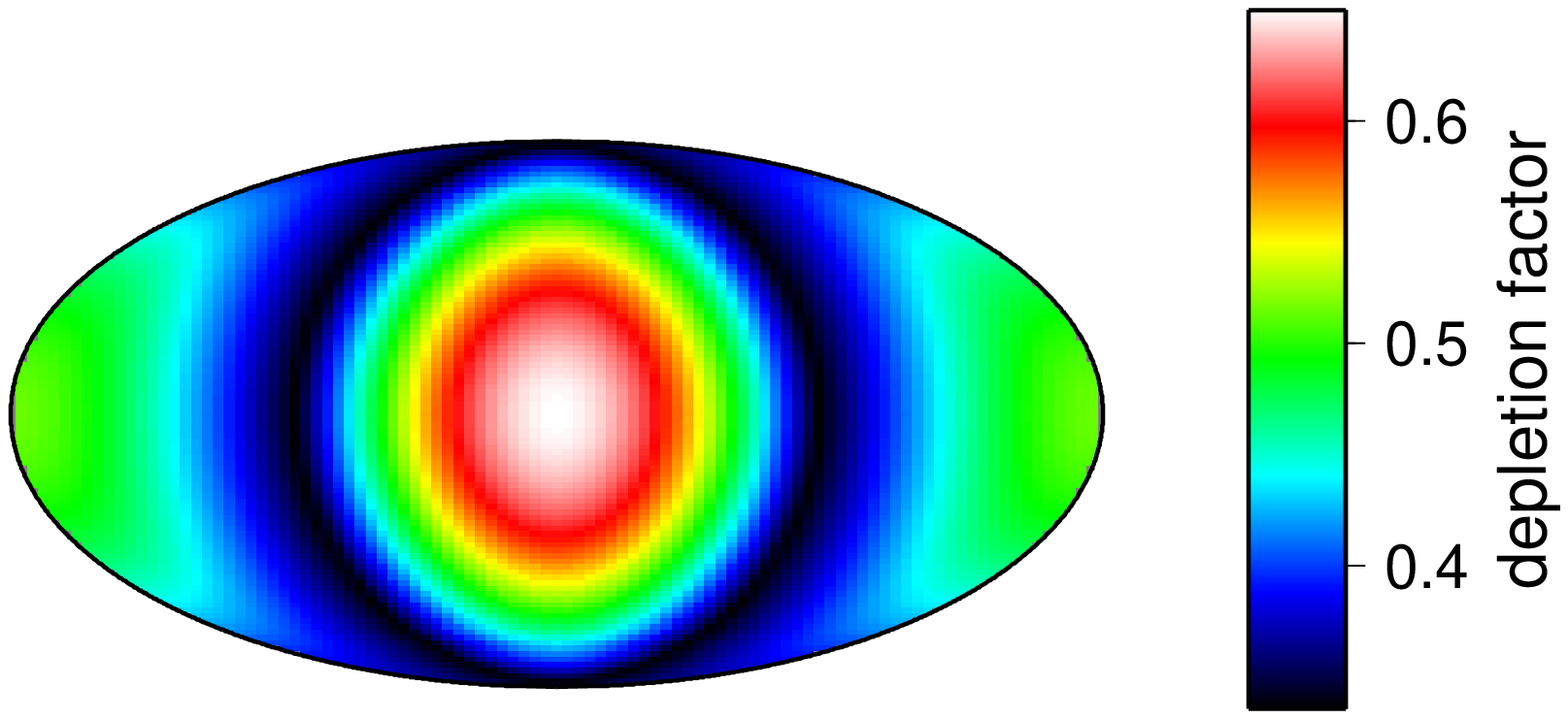}
}
\\
\subfloat[$\varepsilon = 0.15$] {
 \includegraphics[width=0.45\textwidth]{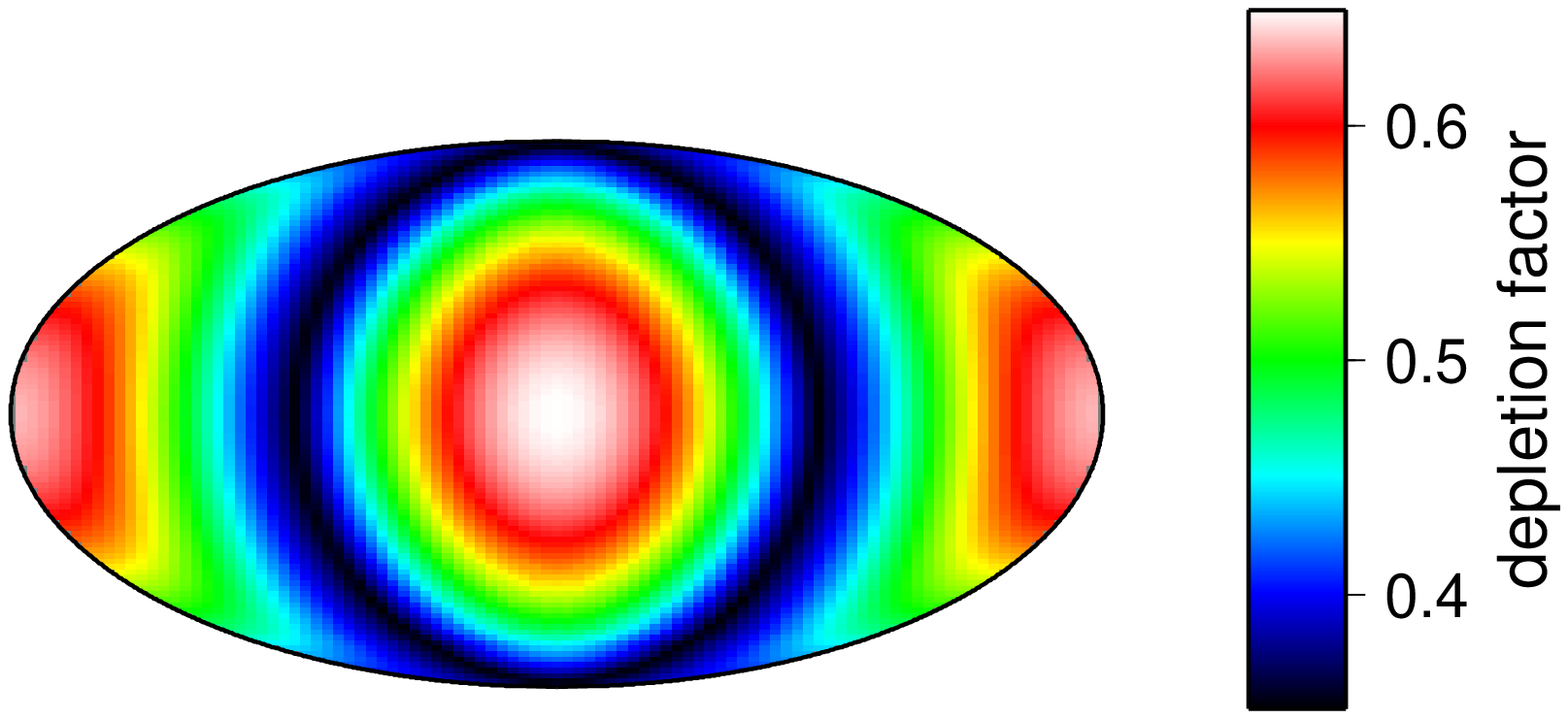}
} 
\end{center}
\caption{ENA depletion maps of the averaged depletion factor $\bar{\epsilon}$ at $E_{obs}$ = 1~keV for different TS geometries. Note that small factors correspond to a large modification and vice versa.}
\label{fig-ena-depletion}
\end{figure*}

Finally, we need to use a model for the TS geometry. We apply
a description where the TS surface has an ellipsoidal form
with two identical short and one longer semi major axes
\citep[as described in e.g.][]{scherer-fahr-09-injection-rate}, where the
numerical eccentricity $\varepsilon = \sqrt{1 - b^2/a^2}$ is used to
characterise the deviation from a perfectly spherical shock (i.e. $\varepsilon = 0$).
Figure \ref{fig-ena-skymap} easily demonstrates that the ENA skymaps
exhibit a significantly different behaviour depending on whether the
depletion correction is used or not. The ring feature that can be identified
as related to the IBEX ribbon emerges in considerably different regions,
and analysing ENA line-of-sight data clearly requires a good handling of
the depletion process.

\section{Depletion maps}

Next, we separate the depletion correction from the rest of the line-of-sight integral
by introducing the effective depletion factor $\bar{\epsilon}$, by rewriting
Eq. \ref{eq-ena-flux} in the form
\begin{equation}
\begin{split}
\Phi_{ENA}(v_{obs}, \alpha ,\delta)
 &= \int_{r_0}^{r_1} dr \cdot j_{p}(E_{ENA}) \epsilon(\phi, \theta, r) \sigma_{ex}(v_{ENA}) n_{H}
\\
 &= \bar{\epsilon} \int_{r_0}^{r_1} dr \cdot j_{p}(E_{ENA}) \sigma_{ex}(v_{ENA}) n_{H},
\end{split}
\label{eq-ena-flux-2}
\end{equation}
or
\begin{equation}
\bar{\epsilon} = \frac{ \int_{r_0}^{r_1} dr \cdot j_{p}(E_{ENA}) \epsilon(\phi, \theta, r) \sigma_{ex}(v_{ENA}) }
{\int_{r_0}^{r_1} dr \cdot j_{p}(E_{ENA}) \sigma_{ex}(v_{ENA}) }.
\label{eq-epsilon}
\end{equation}
This expression strongly depends on the streamline profile due to a frame-of-reference
transformation, where the curved streamlines result in different regions of the phase
space of $j_p(E_{ENA})$ being sampled along a line-of-sight. In addition, the integral
also samples multiple different plasma flowlines, which might be characterized by
qualitatively and quantitatively different source proton distribution functions. Therefore,
the extinction profile calculated with the help of Eq. \ref{eq-epsilon} strongly depends
on model assumptions in the inner heliosheath, which offers an attractive method
to verify or disprove existing numerical and analytical models of this region.

To demonstrate this behaviour, we present extinction maps for a few selected configurations
in Fig. \ref{fig-ena-depletion}. In these figures, we have adopted a source distribution function
following a power law of the form $v^{-5}$, with a constant number density on the
near downstream side of the termination shock. These figures easily demonstrate that the general
shape of the depletion map is almost independent on the geometry of the outer heliosphere,
and that the difference in the ENA skymaps presented on Fig. \ref{fig-ena-skymap}
is primarily due to a suppression of the ENA emission feature that appears in the
extinctionless skymaps. However, the extinction effect is slight weaker than the emission
feature, which allows the ring feature to persist in a certain range of geometric parameters.

One additional point worth mentioning is the extinction factor in the heliotail region,
where a more significant modification of the depletion factor is observed, ranging between
$0.55 \leq \bar{\epsilon} \leq 0.65$ for the presented range of geometry parameters. This
result suggests that the depletion mechanism might provide an additional insight
in studies of the shape of the heliotail, a region that has recently found renewed interest in the
IBEX community \citep{mccomas-13-deflected-tail}.

\section{The depletion correction above 10~keV}

\begin{figure*}[tbp]
\begin{center}
\subfloat[E=10~keV] {
 \includegraphics[width=0.45\textwidth]{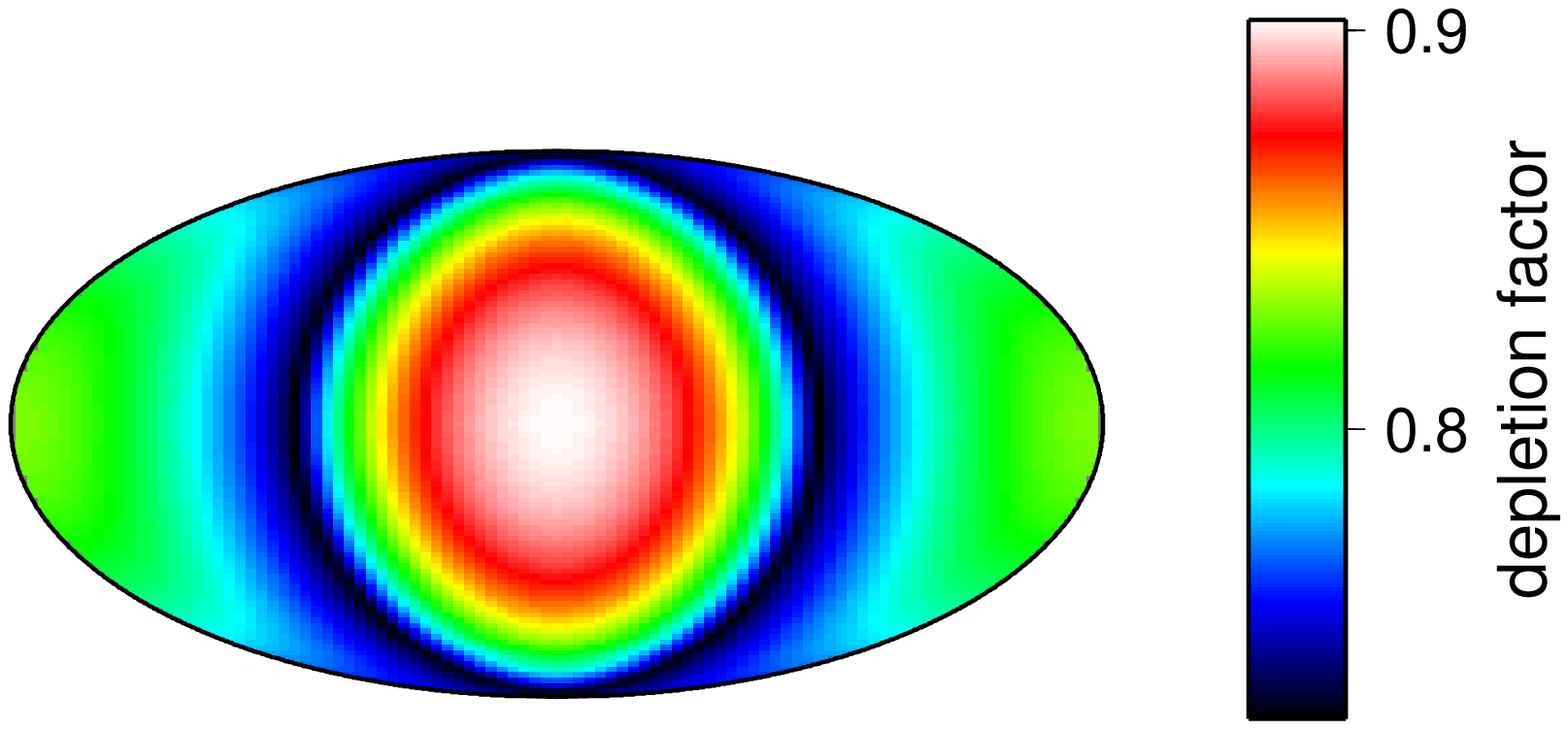}
} 
\hspace{0.05\textwidth}
\subfloat[E=30~keV] {
 \includegraphics[width=0.45\textwidth]{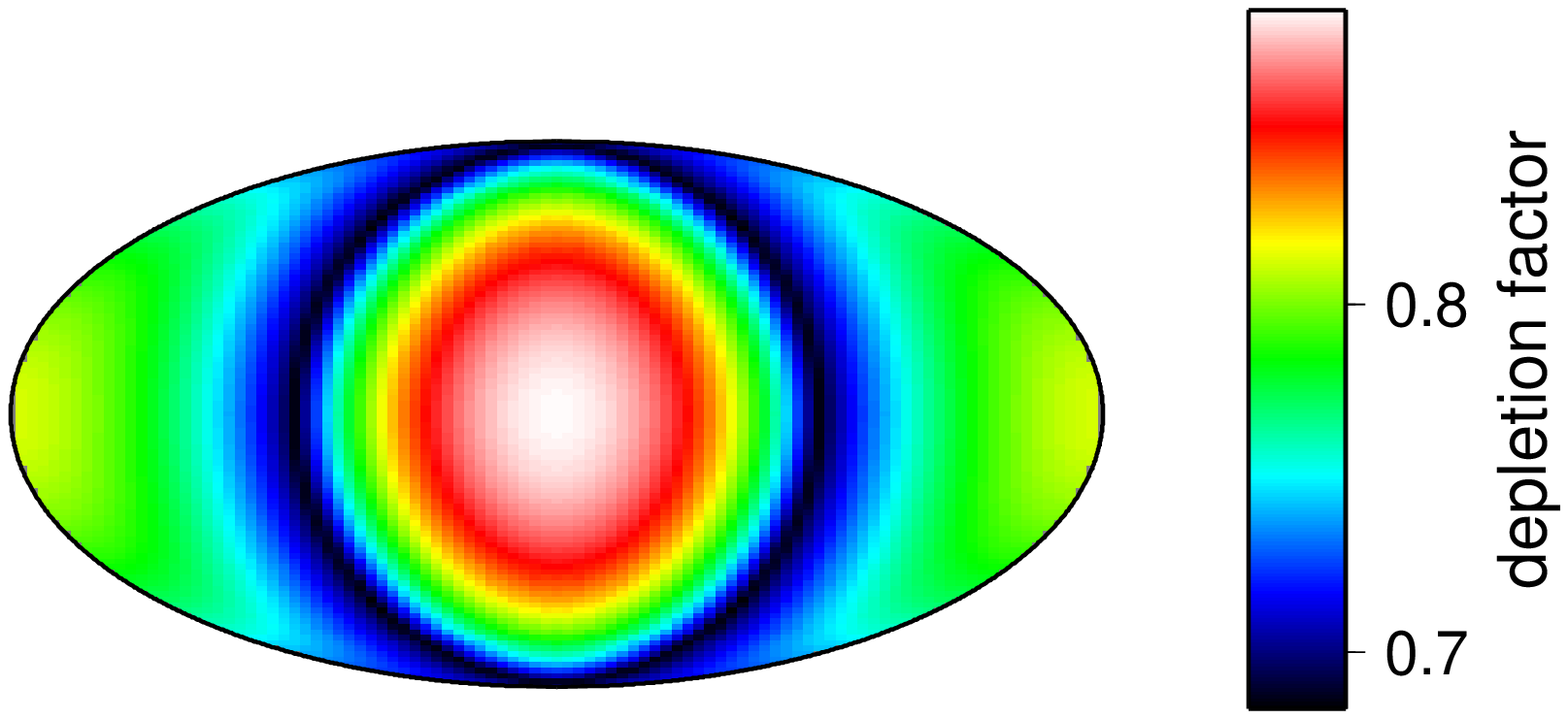}
}
\\
\subfloat[E=50~keV] {
 \includegraphics[width=0.45\textwidth]{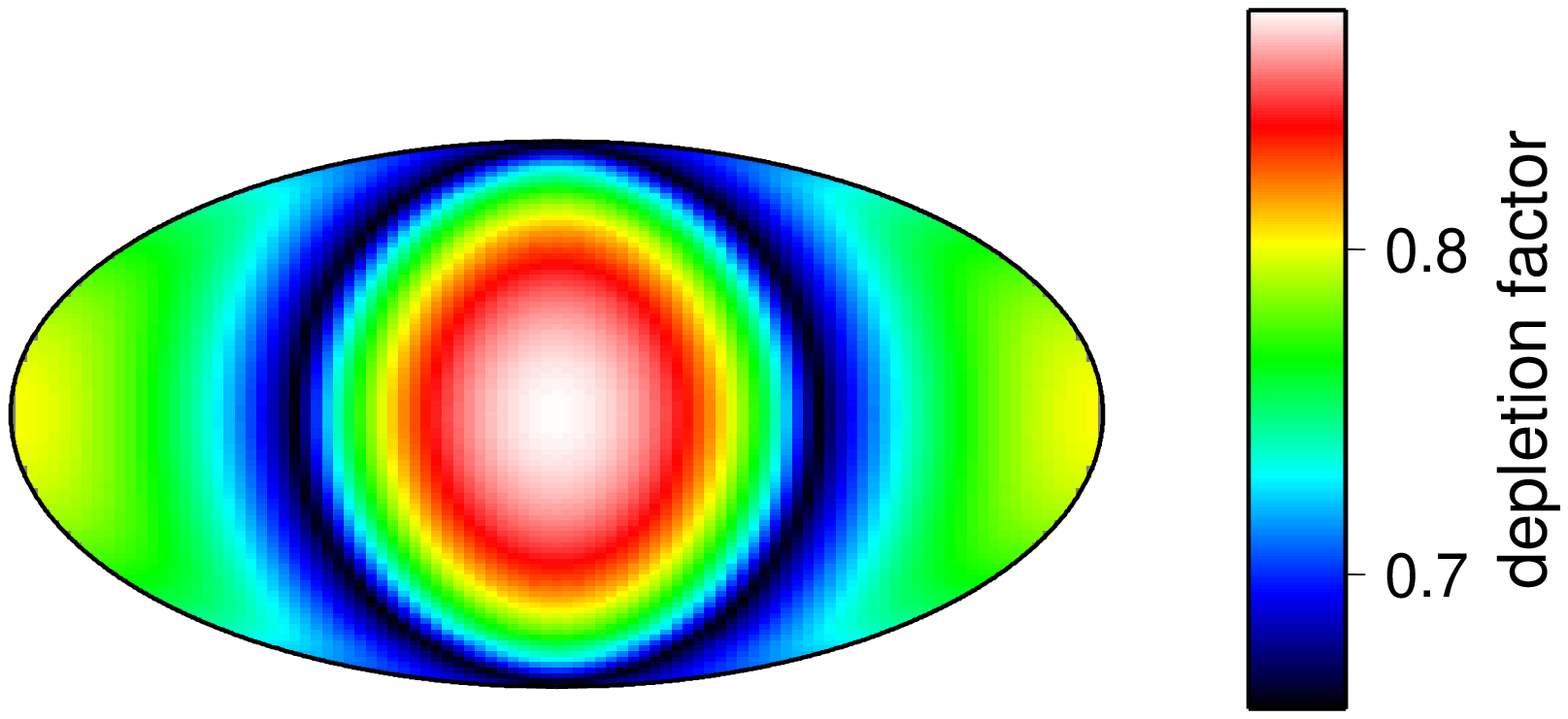}
} 
\end{center}
\caption{ENA depletion maps of the averaged depletion factor $\bar{\epsilon}$ for a numerical eccentricity of $\varepsilon=0.05$ for 10, 30 and 50~keV protons. Note that small factors correspond to a large modification and vice versa.}
\label{fig-energy}
\end{figure*}

In this section, we present first calculations for the depletion correction at higher energies.
INCA/CASSINI observes ENA fluxes in the energy intervall between 5 and 50~keV, while
the upcoming IMAP mission (the follower to the IBEX mission) is supposed to
even include a detector for ENAs until up to 100~keV, and therefore we now study
the depletion correction in the energy range between 10 and 100~keV.

At these higher energies, the product $v_{rel}\cdot\sigma_{ex}$ starts to drop
due to the behaviour of the charge exchange cross section
\citep[see e.g.][]{lindsay-stebbings-05-charge-exchange}, resulting in a
weaker depletion effect, while at the same time, the source distribution function
is assumed to drop like $f(v) \propto v^{-4...-5}$
\citep{fahr-chashei-verscharen-pui-power-laws-09}, resulting in much weaker global
ENA fluxes. On Fig. \ref{fig-energy}, we present depletion factors at higher energies,
demonstrating that the general form of the depletion map is nearly unchanged, with
only the absolute values of $\bar{\epsilon}$ being modified systematically towards higher
values, i.e. a weaker depletion effect. This behaviour demonstrates that the overall form
of the plasma flowline profile is well reflected by the depletion correction at all energies
relevant for ENA observations, almost independent of the energy range in question.

One interesting trend that emerges in Fig. \ref{fig-energy} is that, towards higher
energies, the strongest depletion effect that is primarily found in the flanks of the
heliosphere exhibits a slightly different behaviour than the depletion in the
upwind (nose) and downwind (tail) directions, with the depletion growing slightly stronger with
increasing energies. This is an effect of the streamline orientation; in the flanks,
most streamlines sampled by the line-of-sight integral are almost perpendicular to
said line-of-sight, and therefore, the velocity transformations discussed earlier
result in a different relative velocity between the energetic ions and the cold interstellar
hydrogen that is used to determine the charge exchange cross section.

\section{Conclusions}

The source depletion process identified by \cite{sf13-transit-time}
introduces a new and significant aspect to the interpretation
of IBEX ENA data, which we have now quantified for the first time.
As we have demonstrated, understanding the plasma flowline profile
in the inner heliosheath is critcal for understanding an inner
heliosheath ENA source, as the resulting depletion factors along
the line-of-sight do significantly modify the integrated ENA fluxes.

For this reason, we conclude that a realistic description of the
inner heliosheath is clearly required for a full data analysis.
Unfortunately, this task is greatly complicated by the very unclear
observational data taken by the Voyagers, which suggests that the heliopause,
i.e. the distant boundary of the heliosheath, is a very nonideal
structure dominated by a combination of mostly misunderstood microphysics.
Because of this, there is no clear hint towards which of the various models of the
heliospheric boundary found in the literature would be the most realistic
approximation; in fact, it is probable that our adopted model of this
region of the heliosphere requires a major revision.

Nevertheless, our results clearly prove that the geometry and the
flowline profile of the outer heliosphere leaves a strong imprint on the
observed ENA skymaps, that is almost independent of the ENA energies in question.
Therefore, understanding the outer heliosphere and the source depletion effect
is of high importance for understanding ENAs generated in this region.
Using the same arguments, it should be possible to differentiate between
competing models for the outer heliosphere by calculating the characteristic
depletion skymap ($\bar{\epsilon}$) in a way similar to the one we have presented here.
Finally, our results demonstrate that the depletion effect is globally weaker in
the energy range between 10 and 100~keV, allowing the resulting ENA fluxes to
increase by a factor of 2 compared to fluxes at 1~keV. This doubles the
ENA fluxes to be expected above 10~kev, which suggests that
the ENA fluxes observed at higher energies by INCA/CASSINI and the upcoming IMAP mission
should be higher than expected from a simple extrapolation of energetic ion spectra,
resulting in a harder spectrum at the highest energies.

\begin{acknowledgements}
M. Siewert is grateful to the Deutsche Forschungsgemeinschaft for financial support granted to him in the frame of the project Si-1550/2-2.
\end{acknowledgements}

\bibliographystyle{aa}
\bibliography{aniso,theory-plasma,experiment-plasma,voyager,ibex,bk-phys,my_papers}

\end{document}